\documentclass[twocolumn]{aastex631}

\providecommand{\bjdtdb}{\ensuremath{\rm {BJD_{TDB}}}}
\providecommand{\tjdtdb}{\ensuremath{\rm {TJD_{TDB}}}}

\providecommand{\msun}{\ensuremath{\,M_\Sun}}
\providecommand{\rsun}{\ensuremath{\,R_\Sun}}
\providecommand{\lsun}{\ensuremath{\,L_\Sun}}
\providecommand{\mj}{\ensuremath{\,M_{\rm J}}}

\providecommand{\fave}{\langle F \rangle}
\providecommand{\fluxcgs}{10$^9$ erg s$^{-1}$ cm$^{-2}$}
\usepackage{tabularx}
\usepackage{amsmath}

\shorttitle{No Thick Atmosphere on Gl 486b}
\shortauthors{Weiner Mansfield et al.}

\begin{document}

\title{No Thick Atmosphere on the Terrestrial Exoplanet Gl 486b}

\author[0000-0003-4241-7413]{Megan Weiner Mansfield}
\affiliation{Steward Observatory, University of Arizona, Tucson, AZ, USA}
\affiliation{School of Earth and Space Exploration, Arizona State University, Tempe, AZ, USA}
\affiliation{NHFP Sagan Fellow}

\author[0000-0002-6215-5425]{Qiao Xue}
\affiliation{Department of Astronomy \& Astrophysics, University of Chicago, Chicago, IL, USA}

\author[0000-0002-0659-1783]{Michael Zhang}
\affiliation{Department of Astronomy \& Astrophysics, University of Chicago, Chicago, IL, USA}
\affiliation{51 Pegasi b fellow}

\author[0009-0000-8049-3797]{Alexandra S. Mahajan}
\affiliation{Center for Astrophysics \textbar \ Harvard \& Smithsonian, 60 Garden St, Cambridge, MA, USA}

\author[0000-0003-2775-653X]{Jegug Ih}
\affiliation{Department of Astronomy, University of Maryland, College Park, MD, USA}

\author[0000-0002-9076-6901]{Daniel Koll}
\affiliation{Peking University, Beijing, People's Republic of China}

\author[0000-0003-4733-6532]{Jacob L.\ Bean}
\affiliation{Department of Astronomy \& Astrophysics, University of Chicago, Chicago, IL, USA}

\author[0000-0002-0508-857X]{Brandon Park Coy}
\affiliation{Department of Geophysical Sciences, University of Chicago, Chicago, IL, USA}

\author[0000-0003-3773-5142]{Jason D.\ Eastman}
\affiliation{Center for Astrophysics \textbar \ Harvard \& Smithsonian, 60 Garden St, Cambridge, MA, USA}

\author[0000-0002-1337-9051]{Eliza M.-R. Kempton}
\affiliation{Department of Astronomy, University of Maryland, College Park, MD, USA}

\author[0000-0002-1426-1186]{Edwin S. Kite}
\affiliation{Department of Geophysical Sciences, University of Chicago, Chicago, IL, USA}

\begin{abstract}

A primary science goal for \textit{JWST} is to detect and characterize the atmospheres of terrestrial planets orbiting M dwarfs (M-Earths). The existence of atmospheres on M-Earths is highly uncertain because their host stars' extended history of high XUV irradiation may act to completely remove their atmospheres. We present two \textit{JWST} secondary eclipse observations of the M-Earth Gl~486b (also known as GJ~486b) between $5-12$~$\mu$m. We combined these observations with a precise analysis of the host star parameters to derive a planetary dayside temperature of $T_{p}=865 \pm 14$~K. We compared this temperature to the maximum expected temperature for a zero albedo, zero heat redistribution bare rock and derived a temperature ratio of $\mathcal{R}=\frac{T_{p,dayside}}{T_{p,max}}=0.97 \pm 0.01$. This value is consistent with an airless body with a slight non-zero albedo or a thin atmosphere with $<1$\% H$_{2}$O or $<1$~ppm CO$_{2}$. However, it is inconsistent with an Earth- or Venus-like atmosphere, and the spectrum shows no clear emission or absorption features. Additionally, our observations are inconsistent with the water-rich atmospheric scenario allowed by previous transit observations and suggest the transmission spectrum was instead shaped by stellar contamination \citep{Moran2023}. Given the potential for atmospheric escape throughout the system's $\geq6.6$-Gyr lifetime \citep{DiamondLowe2024}, we conclude that the observations are likely best explained by an airless planet. This result is the most precise measurement yet of terrestrial exoplanet thermal emission with \textit{JWST}, which places a strong constraint on the position of the ``Cosmic Shoreline'' between airless bodies and those with atmospheres.


\end{abstract}


\section{Introduction}
\label{sec:intro}

A key goal of \textit{JWST} is to identify and characterize atmospheres on terrestrial planets. In particular, \textit{JWST} is the first facility capable of searching for atmospheres on a large sample of rocky planets orbiting M dwarfs (M-Earths). However, the ability of M-Earths to retain atmospheres over geologic timescales is highly uncertain. The high X-ray and ultraviolet flux (XUV) of young M dwarfs  \citep[e.g.,][]{Tian2009,Davenport2012}, and their high flare rates at older ages \citep{Johnstone2020} may strip terrestrial planets in close-in orbits of their atmospheres.

Studies of solar system bodies have suggested that atmospheric loss or retention is determined by a body's position relative to the ``Cosmic Shoreline'', a dividing line which separates bodies at low bolometric or XUV fluxes which are able to retain atmospheres from the airless bodies at high bolometric/XUV flux, as a function of surface gravity \citep{Zahnle2017}. Extending this theory to M-Earths a priori is complicated by several unknown factors, such as the variation in XUV flux among similar stars \citep{West2008}; and the effects of atomic line cooling \citep{Nakayama2022}, initial volatile inventory \citep{Kite2020}, and the timing of volatile delivery \citep{Dorn2018} or outgassing \citep{Tian2009} on the prevalence of atmospheres on M-Earths. Therefore, a first order goal of \textit{JWST} exoplanet science is to precisely measure the presence or absence of atmospheres on M-Earths to observationally constrain the location of this ``Cosmic Shoreline''.

Gl 486b, also known as GJ 486b, is a terrestrial planet ($R_{p}=1.289^{+0.019}_{-0.014}R_{\Earth}$, $M_{p}=2.770^{+0.076}_{-0.073}M_{\Earth}$) recently discovered by TESS orbiting an M3.5 dwarf \citep{Trifonov2021}. At a distance of $8.0756$~pc , it is the third-closest known transiting exoplanet system to host a terrestrial exoplanet, and its high emission spectroscopy metric \citep[$\mathrm{ESM}=21$,][]{Kempton2018} makes it a prime target for \textit{JWST} observations. Recent X-ray and UV observations of the host star Gl 486 indicate that Gl 486b may have been able to retain an atmosphere with a reasonable initial volatile inventory \citep{DiamondLowe2024}. However, despite appearing relatively inactive at optical wavelengths \citep{Caballero2022}, it was also observed to have high-energy XUV flares \citep{DiamondLowe2024}. These recent observations also used an age-activity relationship to place a lower limit of 6.6~Gyr on the age of the host star Gl~486.

The transmission spectrum of Gl 486b was observed with \textit{JWST} NIRSpec/G395H between $3-5$~$\mu$m \citep{Moran2023}. However, these observations were unable to definitively determine whether the planet had an atmosphere, as the transmission spectrum was consistent with either a water-rich atmosphere or an airless body with a spectrum affected by stellar contamination. 

In this paper we present the secondary eclipse spectrum of Gl~486b observed using \textit{JWST}'s Mid-Infrared Instrument (MIRI). We describe our observations and data reduction in Section~\ref{sec:obs}, including an analysis of archival data to precisely constrain the stellar parameters. We describe the calculation of the observed broadband dayside temperature and compare it to expectations for atmospheres and airless bodies in Section~\ref{sec:daytemp}. In Section~\ref{sec:model}, we compare the resulting spectra to atmosphere forward models and discuss our result in the context of other observations of M-Earths near the Cosmic Shoreline. Finally, we conclude in Section~\ref{sec:conclude}.

\section{Observations and Data Reduction}
\label{sec:obs}

We observed two secondary eclipses of Gl~486b with \textit{JWST}/MIRI using the Low Resolution Spectroscopy \citep[LRS,][]{Kendrew2015} Time Series Observation mode, which covers wavelengths from $5-12$~$\mu$m. The observations were taken on May 29 and June 1, 2023, as part of program GO 1743 (PI Mansfield)\footnote{All of the data presented in this paper were obtained from the Mikulski Archive for Space Telescopes (MAST) at the Space Telescope Science Institute. The specific observations analyzed can be accessed via \url{http://dx.doi.org/10.17909/3tqz-y166}. The data behind the figures can be found on Zenodo \citep{ZenodoData}.}. Each observation consisted of a single 4.6-hr exposure with 14,810 integrations and 6 groups/integration. The peak pixel counts were $\approx$58,600~DN ($\approx86$\% full well).

\begin{figure}
    \centering
    \includegraphics[width=\linewidth]{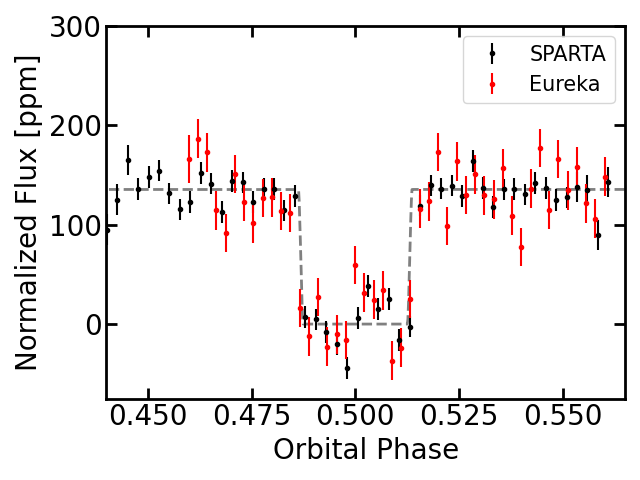}
    \caption{White light curves for both the \texttt{SPARTA} (black) and \texttt{Eureka!} (red) reductions of observations of Gl~486b. The dashed gray line shows the best fit eclipse depth from the \texttt{SPARTA} reduction.}
    \label{fig:white}
\end{figure}

\begin{figure}
    \centering
    \includegraphics[width=\linewidth]{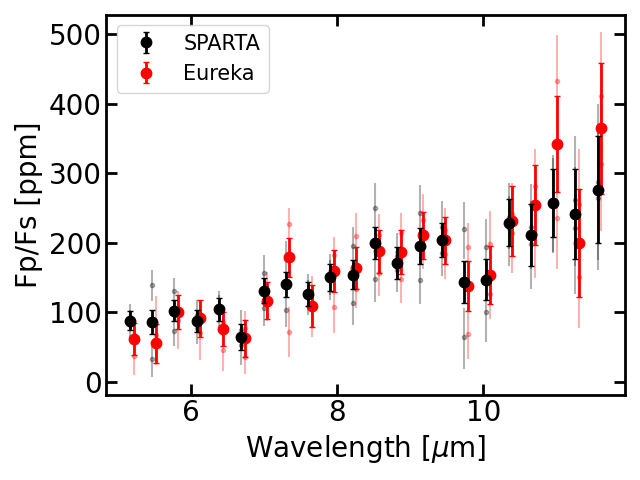}
    \caption{MIRI/LRS eclipse spectra of Gl~486b from the reductions with \texttt{SPARTA} (black) and \texttt{Eureka!} (red). Thin points show results from the two individual eclipses, and thick points show weighted averages of both eclipses combined. The two data reductions are offset slightly in wavelength for clarity. Both independent reductions of both eclipses are in good agreement with each other.}
    \label{fig:spec}
\end{figure}

We applied two independent data reduction pipelines, \texttt{SPARTA} and \texttt{Eureka!}, to ensure reproducible results. 
Figures~\ref{fig:white} and \ref{fig:spec} and Table~\ref{tab:spectra} shows resulting eclipse spectra. The two methods showed consistent white light and spectroscopic eclipse depths within $1\sigma$.

\begin{table}
\centering
\begin{tabularx}{\linewidth}{>{\centering\arraybackslash}X|>{\centering\arraybackslash}X >{\centering\arraybackslash}X}
    Wavelength Range [$\mu$m] & \texttt{SPARTA} Eclipse Depth [ppm] & \texttt{Eureka!} Eclipse Depth [ppm] \\
    \hline
    $5.06-5.365$ & $88 \pm 14$ & $62 \pm 23$ \\
    $5.365 - 5.67$ & $86 \pm 17$ & $55 \pm 28$ \\
    $5.67 - 5.975$ & $102 \pm 15$ & $100 \pm 25$ \\
    $5.975 - 6.28$ & $87 \pm 16$ & $91 \pm 27$ \\
    $6.28 - 6.585$ & $104 \pm 17$ & $76 \pm 25$ \\
    $6.585 - 6.89$ & $64 \pm 19$ & $62 \pm 26$ \\
    $6.89 - 7.195$ & $131 \pm 18$ & $117 \pm 27$ \\
    $7.195 - 7.5$ & $140 \pm 18$ & $179 \pm 28$ \\
    $7.5 - 7.805$ & $126 \pm 17$ & $109 \pm 30$ \\
    $7.805 - 8.11$ & $150 \pm 20$ & $160 \pm 30$ \\
    $8.11 - 8.415$ & $154 \pm 21$ & $163 \pm 30$ \\
    $8.415 - 8.72$ & $199 \pm 23$ & $188 \pm 31$ \\
    $8.72 - 9.025$ & $171 \pm 23$ & $187 \pm 32$ \\
    $9.025 - 9.33$ & $195 \pm 26$ & $210 \pm 34$ \\
    $9.33 - 9.635$ & $204 \pm 25$ & $203 \pm 34$ \\
    $9.635 - 9.94$ & $143 \pm 30$ & $138 \pm 36$ \\
    $9.94 - 10.245$ & $147 \pm 30$ & $153 \pm 42$ \\
    $10.245 - 10.55$ & $229 \pm 34$ & $231 \pm 50$ \\
    $10.55 - 10.855$ & $211 \pm 44$ & $254 \pm 58$ \\
    $10.855 - 11.16$ & $257 \pm 49$ & $343 \pm 69$ \\
    $11.16 - 11.465$ & $241 \pm 65$ & $200 \pm 77$ \\
    $11.465 - 11.77$ & $276 \pm 77$ & $365 \pm 94$ \\
    \hline
    $5.06-11.77$ & $135.5 \pm 4.9$ & $133.7 \pm 6.6$ \\
\end{tabularx}
\caption{\label{tab:spectra} Secondary eclipse spectra from both data reductions. The \texttt{SPARTA} reduction is the primary one used for analysis, while the \texttt{Eureka!} one is included for completeness. The last line lists the broadband eclipse depths.}
\end{table}

\subsection{SPARTA}
\label{sec:sparta}

SPARTA is a completely independent JWST data reduction package that does not use any code from any other pipeline, including the standard JWST pipeline.  It was first introduced in \cite{Kempton2023}, and subsequently used in several data analyses (e.g. \citealt{Bell2022,zhang_2024}).  This paper most closely follows the methodology in \cite{zhang_2024} (see their Appendix A).  A brief summary follows.

First, we perform nonlinearity correction, dark subtraction, multiplication by the gain (assumed to be 3.1 electrons/DN), two rounds of up-the-ramp fitting to take out uncorrected nonlinearities, and flat fielding.  The two rounds of fitting are meant to take out uncorrected non-linearities: we discard the last group in the first round of fitting, obtain the median residuals across all integrations, add the residuals to the original data, and repeat the up-the-ramp fit.

Afterwards, we remove the background at each wavelength (corresponding to a column, since we rotate the image 90 degrees) by subtracting the median of rows 10:25 and -25:-10 (Python indexing convention) for each column.

Using these background-removed images, we calculate a median image across all integrations, and use it as a template to calculate the position offset of the trace in each integration in both the x and y directions.

Using the previously calculated median image to generate a spatial profile, we use optimal extraction to extract a spectrum for each integration.  We use an extraction half-width of 5 pixels, and reject pixels more than 5$\sigma$ away from the model image as outliers.

The spectra for each integration are gathered into one pickle file.  Integrations that are $>4\sigma$ outliers in the white light curve are rejected, while $>4\sigma$ outliers in the unbinned spectroscopic light curves are repaired via interpolation.  

Finally, we use the Markov chain Monte Carlo (MCMC) package \texttt{emcee} to fit the white light curve (5--12 $\mu$m) with an eclipse model and a systematics model after trimming the first 30 minutes (1616 integrations) of data.  The free parameters are the eclipse time, eclipse depth, normalization factor, exponential ramp parameters (amplitude and timescale), slope with time, linear detrending parameters with x and y position, and an error inflation factor.  The period, Rp/Rs, a/Rs, and inclination are fixed to the ``weighted mean'' values from the FIREFLy reduction of NIRSpec/G395H transit data \citep{Moran2023}.  The two eclipse times were $60093.3530 \pm 0.0002$ and $60096.2874 \pm 0.0002$ (BMJD), which are $12 \pm 15$ s and $22 \pm 15$ s later than the transit time plus half the period -- fully consistent with zero eccentricity when the 17 s light travel time is subtracted out.

To fit the spectroscopic light curves, we fix the eclipse time to $60093.3530$ (BMJD\textsubscript{TDB}) for the first visit and to that value plus twice the period for the second visit.  Other system parameters were also fixed to the same values as for the white light curve.  The systematics model is identical to the one we used for the white light curve: a linear trend with time, x, and y, multiplied by an exponential ramp, all multiplied by a normalization factor.  We use \texttt{emcee} \citep{ForemanMackey2013} to simultaneously fit for the eclipse depth, the systematics parameters, and the error inflation for each wavelength bin, after trimming the first 30 minutes of data.  We divide the bandpass into 22 wavelength bins, spanning 5.060 to 11.770 $\mu$m in 0.305 $\mu$m intervals.  This binning scheme was chosen to mitigate the ``shadowed region'' effect (seen in visit 2, but not evident in visit 1), because the reddest 4 bins (between 10.55 and 11.77 $\mu$m) perfectly span this region.

\subsection{Eureka!}
\label{sec:eureka}

The second data reduction was performed with version 0.10 of the \texttt{Eureka!}\ pipeline \citep{Bell2022}. \texttt{Eureka!}\ is divided into five stages. Stages 1 and 2 use the standard \texttt{JWST} pipeline to convert from ramp data to slopes and perform calibrations such as flat fielding and unit conversion. Stage 3 extracts spectra from 2D images, Stage 4 creates binned spectroscopic light curves, and Stage 5 fits the data with a \texttt{batman} model for astrophysical flux \citep{batman} and systematic models.

We used the standard \texttt{Eureka!}\ set-up described in \citet{Bell2022}, except for the following changes. In Stage 1, we skipped the jump step, as it can introduce noise for observations with a small number of groups. We also assumed a gain of 3.1 electrons/DN, which has been shown to be consistent with on-sky observations \citep{Kempton2023}. In Stage 2, we skipped the photometric flux calibration step to improve estimation of uncertainties. In Stage 3, we extracted the spectrum via optimal extraction \citep{Horne1986} from a window with a half-width of 6 pixels centered on the location of the spectral trace. We rotated the MIRI spectra so that rows were along the dispersion direction and subtracted the background by fitting a constant column-by-column to the region more than 14 pixels away from the spectral trace on either side. We used a $5\sigma$ threshold for outlier rejection during background subtraction and a $10\sigma$ threshold while fitting the median image spatial profile. In Stage 4, we corrected for drift along the direction of the spectral trace, as discussed in more detail below. We also performed $10\sigma$ sigma clipping on the binned 1D time series.

In Stage 5, we fit the data with a model for the secondary eclipse and a visit-long polynomial systematics trend. We clipped the first 3234 integrations from the beginning of each observation to remove a visible ramp. For the observations taken on May 29 and June 1, we used a zeroth-order and first-order polynomial, respectively. For the June 1 eclipse, the planet-to-star radius ratio ($r_{p}/r_{s}$), period ($p$), inclination ($inc$), and eccentricity ($ecc$) were fixed to the best-fit values from the joint stellar and planetary parameter fit described in Section~\ref{sec:star}. The mid-eclipse time ($T_{mid}$) and semi-major axis to stellar radius ratio ($a/r_{s}$) were free parameters in the white light curve fit then fixed to the best-fit value for the spectroscopic light curve fits. The eclipse depth ($F_{p}/F_{s}$), polynomial coefficients ($c_{0}$ and $c_{1}$), and a scaling parameter to account for a different level of noise than the Stage 3 prediction ($mult$) were allowed to vary freely in both the white light and spectroscopic fits. The May 29 observation had a higher level of scatter, as described below, so for this fit we fixed $a/r_{s}$ and $T_{mid}$ to the best-fit values from the June 1 white light curve. The other parameters were treated in the same manner as for the June 1 observation. We estimated the parameters with an MCMC fit using the \texttt{emcee} package \citep{ForemanMackey2013}, with 200 walkers, 3500 steps, and a 350-step burn-in.

The observations taken on May 29 showed greater drift in the position of the spectral trace. We first attempted to correct for this by adding free parameters to decorrelate against the x and y positions and widths of the spectral trace, similar to the \texttt{SPARTA} data reduction. However, this decorrelation was unsuccessful in removing the drift noise and left behind approximately hour-long changes in the measured flux of $\approx100$~ppm, at a similar level as the secondary eclipse itself. We additionally tested decorrelating against the x and y positions recorded from \texttt{SPARTA}, which were not identical to the x and y positions recorded by \texttt{Eureka!}, but found that this also left structured noise on the order of 100 ppm. We instead opted to use the drift correction built into Stage 4 of \texttt{Eureka!}, which cross-correlates spectra to determine the relative drift of each integration. We used a 5 pixel half-width, although we found similar results with many different parameters for the drift correction. In all cases, using \texttt{Eureka!}'s drift correction removed the correlated noise but inflated the scatter in the fluxes such that the light curve depths had uncertainties a factor of $\approx5.5$ larger than those measured during the June 1 visit. The \texttt{Eureka!} spectrum obtained from averaging both visits therefore had larger error bars than the \texttt{SPARTA} analysis but, as shown in Figures~\ref{fig:white} and \ref{fig:spec}, measured consistent eclipse depths to within $1\sigma$.

\subsection{Stellar Parameter Estimation}
\label{sec:star}

Correctly translating a measured secondary eclipse depth into a planetary temperature relies on precise knowledge of stellar and planetary parameters such as the stellar effective temperature and semi-major axis. Additionally, expected values of the derived planetary temperature for an airless body or a body with an atmosphere are determined from a global energy balance \citep{Koll2019}, so understanding whether the derived temperature is consistent with an atmosphere also requires precise stellar parameters. Therefore, we used \texttt{EXOFASTv2} \citep{Eastman2019}, following the procedure of \citet{Mahajan2024} to perform a joint fit of several different observations of Gl~486b and precisely constrain stellar and planetary parameters. We simultaneously fit the white light curves from the two secondary eclipses presented in this study, white light curves from two transits taken with \textit{JWST}/NIRSpec \citep[][with separate fits for the NRS1 and NRS2 detectors]{Moran2023}, and data from Sector 23 (March - April 2020) of the Transiting Exoplanet Survey Satellite. We used data from Sector 23 because it was the earliest available observation of Gl~486b with TESS, which maximized the time baseline and therefore provided the tightest constraint on the ephemeris.

As was done in \citet{Xue2024}, we modeled ramps for the \textit{JWST} data. The best-fit parameters are included in Appendix~\ref{append:star} in Table~\ref{tab:GJ486_stellar}. Using the precisely measured stellar density, the \citet{Mann2019} stellar mass, and the bolometric flux from the spectral energy distribution, the joint fit increased the precision on the stellar radius and stellar effective temperature by $3.5\times$ and $2\times$, respectively, compared to the previous best fit \citep{Caballero2022}. Additionally, both of those parameters, as well as the stellar mass, $r_{p}/r_{s}$, and $a/r_{s}$, were consistent within $1\sigma$ with previous fits \citep{Caballero2022,Moran2023}. While the previous fit assumed a circular orbit with zero eccentricity, our fit was able to precisely constrain the eccentricity to $0.00086^{+0.0016}_{-0.00043}$, which is consistent with zero given the Lucy-Sweeney bias on the eccentricity, which is a positive definite quantity \citep{Lucy1971}. The MIRI/LRS white light eclipse depth obtained from the \texttt{EXOFASTv2} fit is $133.3^{+5.0}_{-4.9}$~ppm, which is consistent to within $1\sigma$ with the white light depths derived from the \texttt{SPARTA} and \texttt{Eureka!} data reductions.


\section{Dayside Temperature Calculation}
\label{sec:daytemp}

We follow the method outlined in \citet{Xue2024} to convert the white light eclipse depth from the \texttt{EXOFASTv2} fit to a planetary dayside temperature. Briefly, we took the planet-to-star flux ratio ($F_{p}/F_{s}$), planet-to-star radius ratio ($R_{p}/R_{s}$), stellar effective temperature ($T_{s}$), stellar metallicity ($[\mathrm{M}/\mathrm{H}]$), and stellar gravity ($\log g$) from each sample in the Markov Chain Monte Carlo fit performed by \texttt{EXOFASTv2}. We then obtained a stellar spectrum by interpolating from a PHOENIX grid \citep{Allard2012} to the correct $T_{s}$, $[\mathrm{M}/\mathrm{H}]$, and $\log g$ using the python package \texttt{pysynphot} \citep{pysynphot}. We tested performing the same fit with an interpolated SPHINX model \citep{Iyer2023} and found consistent results. We convolved the stellar spectrum with a Gaussian kernel to broaden it to the resolution of MIRI/LRS and used it to calculate the photon flux from the planet in each sample following the equation
\begin{equation}
    \label{temp_eq}
    \begin{split}
\frac{F_p}{F_s} 
& = \left(\frac{R_p}{R_s}\right)^2 \cdot \frac{\int \frac{\pi \cdot B_p(T_p, \lambda)}{hc/\lambda} \cdot W_{\lambda}\, d\lambda}{\int \frac{M_{s}(T_s,\,\mathrm{log}\,g,\,[M/H],\,\lambda)}{hc/\lambda}\cdot W_{\lambda} \, d\lambda},
\end{split}
\end{equation}
where $B_p(T_p, \lambda)$ is the planet flux (assuming it acts as a blackbody), $M_{s}(T_s, \log g, [\mathrm{M}/\mathrm{H}], \lambda)$ is the interpolated stellar spectrum, and $W_\lambda$ is the MIRI/LRS throughput. We then invert Equation~\ref{temp_eq} to convert the photon flux to a planetary temperature for each sample.

\begin{figure}
    \centering
    \includegraphics[width=\linewidth]{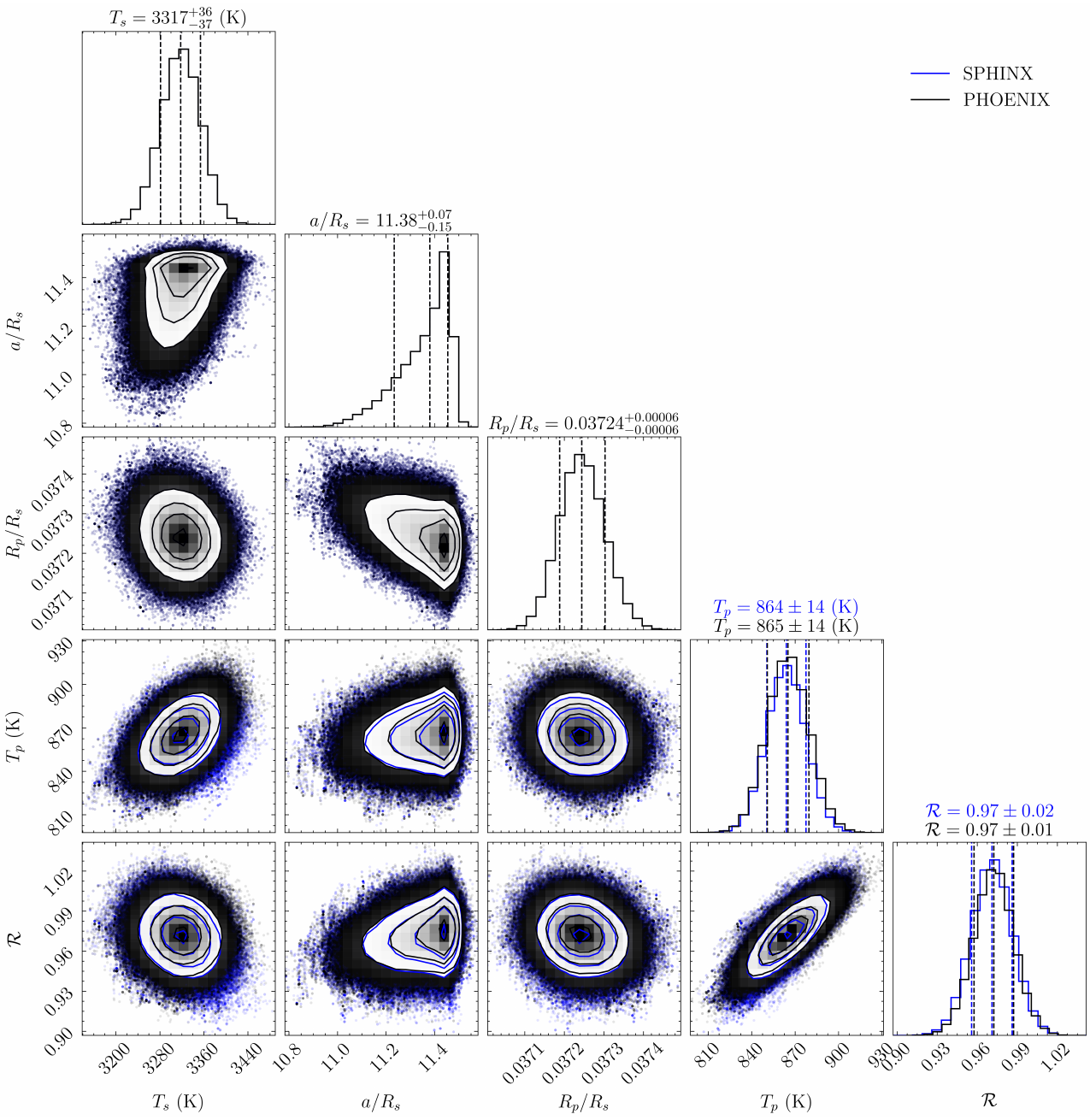}
    \caption{Pairs plot for the best-fit parameters from the joint stellar and planetary fit described in Section~\ref{sec:star}, as well as the derived temperature ratio ($\mathcal{R}$) described in Section~\ref{sec:daytemp}. Black and blue plots show results using a PHOENIX \citep{Allard2012} and SPHINX \citep{Iyer2023} stellar model, respectively. Off-diagonal plots show 2D posterior probabilities for pairs of parameters, with dark, medium, and light shading showing 1, 2, and $3\sigma$ confidence intervals. On-diagonal plots show marginalized posterior probability distributions for each individual parameter.}
    \label{fig:corner}
\end{figure}

Figure~\ref{fig:corner} shows a pairs plot for the stellar parameters and dayside temperature. We derive a planetary dayside temperature of $T_{p}=865 \pm 14$~K. 
This calculation assumes that the MIRI/LRS eclipse depth is a good estimate for bolometric flux, and thus that the planet roughly emits as a blackbody. We make these assumptions because a planet at the temperature of Gl~486b will emit $\approx45$\% of its flux in the LRS bandpass and realistic surfaces emit close enough to a blackbody that this estimate approximates the true dayside temperature well. For example, if the true albedo were 0.1 higher than the albedo we infer from this infrared measurement \citep[a typical difference between the inferred and true bond albedo,][]{Mansfield2019,Whittaker2022}, our dayside temperature would only change by 15~K, or about $1\sigma$.

We neglect thermal beaming, which may result from (e.g.) surface roughness effects \citep{spencer_1990}, and which can raise the observed low-phase-angle brightness temperature of atmosphere-free bodies above $T_{p,max}$ \citep{emery_1988}. This does not affect our conclusion that the atmosphere is probably very thin or absent.

The dayside temperature will depend on both heat redistribution and albedo \citep{Cowan2011}, and the derived joint constraint can be compared to expected values for the heat redistribution and albedo for atmospheres to determine the likelihood that Gl~486b has an atmosphere \citep{Koll2019,Mansfield2019}. Following \citet{Xue2024}, we calculate the temperature ratio $\mathcal{R}$, which compares the derived dayside temperature to the maximum possible temperature for a zero-albedo, zero heat redistribution, smooth bare rock ($T_{p,max}$). We calculate $\mathcal{R}$ using the equation 
\begin{equation}
\begin{split}
     \mathcal{R} &=\frac{T_{p,dayside}}{T_{p,max}} \\
    &= \frac{T_{p,dayside}\sqrt{a/R_s}}{ \left( \frac{2}{3} \right)^{\frac{1}{4}} T_s} ,
\end{split}
\end{equation}
where $a$ is the semi-major axis and $R_{s}$ is the stellar radius. We derive a value of $\mathcal{R}=0.97 \pm 0.01$.

We determine which albedo and heat redistribution values are consistent with the derived $\mathcal{R}$ using the equation
\begin{equation}
    \mathcal{R} =  \left( \frac{2}{3} \right)^{-\frac{1}{4}} \cdot (1-A_B)^{\frac{1}{4}}\cdot(\frac{2}{3}-\frac{5}{12}\varepsilon)^{\frac{1}{4}},
\end{equation}
where $A_{B}$ is the Bond albedo, $\varepsilon$ is the heat redistribution efficiency, and the factor of $\left(\frac{2}{3}\right)^{-1/4}$ is a normalization to ensure that $\mathcal{R}=1$ for the case where there is zero albedo and zero heat redistribution \citep[$A_{B} = \varepsilon=0$,][]{Cowan2011}.

\begin{figure}
    \centering
    \includegraphics[width=\linewidth]{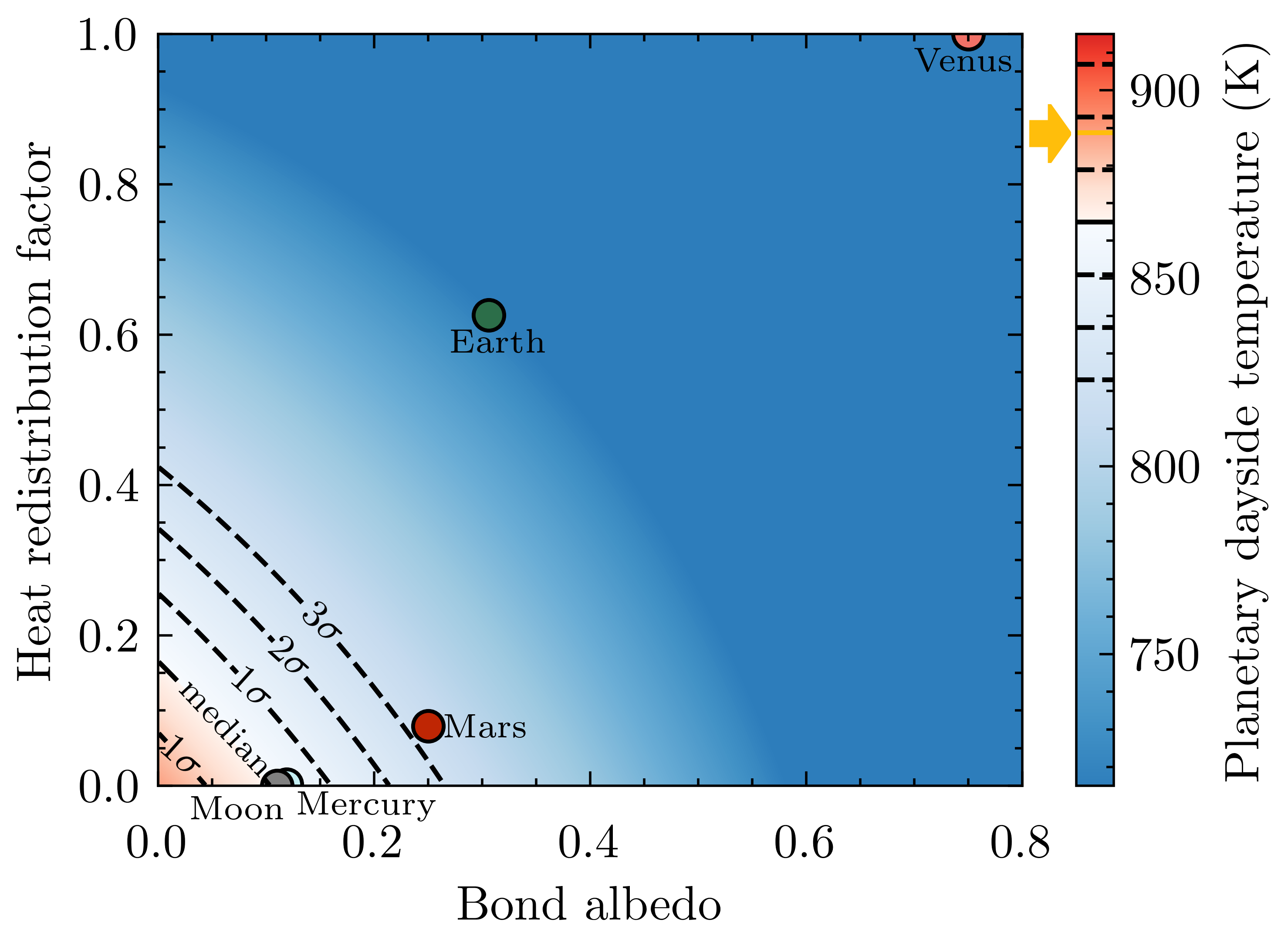}
    \caption{Contour plot showing the planetary dayside temperature of Gl~486b as a function of Bond albedo ($A_{B}$) and heat redistribution factor ($\varepsilon$). Note that T$_\mathrm{s}$, a/R$_\mathrm{s}$ and R$_\mathrm{p}$/R$_\mathrm{s}$ are identical. They are drawn from each sample in the \texttt{EXOFASTv2} MCMC fitting. The solid line shows a contour of the median planetary temperature inferred for Gl~486b, and the surrounding dotted lines show $1\sigma$, $2\sigma$, and $3\sigma$ intervals. The yellow arrow on the colorbar shows the maximum dayside temperature predicted for Gl~486b if it had zero Bond albedo and no heat redistribution. Gray, light blue, red, green, and yellow dots show approximate values for the solar system bodies Moon, Mercury, Mars, Earth, and Venus, respectively. Our observations of Gl~486b are most consistent with airless bodies like the Moon and Mercury that have slight non-zero albedos, and it is highly inconsistent with a thick Earth-like or Venus-like atmosphere.}
    \label{fig:heatmap}
\end{figure}

Figure~\ref{fig:heatmap} shows the joint constraint on albedo and heat redistribution obtained from the dayside brightness temperature of Gl~486b. Figure~\ref{fig:heatmap} compares this constraint with values for the solar system terrestrial planets and the Moon. We assume albedos of $A_{B}=0.11$, 0.119, 0.75, 0.306, and 0.250 for the Moon, Mercury, Venus, Earth, and Mars, respectively. The primary global temperature difference for the fast-rotating solar system bodies is the equator-to-pole temperature difference, rather than the dayside-to-nightside temperature difference that is used to calculate the heat redistribution factor for slow-rotating hot exoplanets. Therefore, following \citet{Cowan2011}, we derive the heat redistribution factor for the solar system bodies using the equation
\begin{equation}
    \varepsilon = \frac{8/3}{(T_{equator}/T_{pole})^{4}+5/3},
\end{equation}
where $T_{equator}$ and $T_{pole}$ are the mean equator and pole temperatures, respectively.

We find that our observations of Gl~486b are most consistent with airless bodies with small non-zero albedos of $A_{B} \approx 0.1$, similar to the Moon and Mercury. Although the observations are consistent to within $\approx3\sigma$ with the presence of a thin atmosphere such as that found on Mars, they are highly inconsistent with thick, Earth- or Venus-like atmospheres.

\section{Comparison to Forward Models}
\label{sec:model}

Although our results show that the data are strongly inconsistent with the presence of a thick atmosphere, we compared a set of forward models to our data to determine which types of atmospheres would still be allowable. We followed the methods of \citet{Whittaker2022} and \citet{Ih2023} to use \texttt{HELIOS} to construct forward modeled thermal emission spectra for Gl~486b. We used a SPHINX stellar model \citep[][updated May 30th, 2024]{Iyer2023} for the input stellar flux. We modeled atmospheres with surface pressures of $10^{-4}-100$~bar. We modeled atmospheres with pure H$_{2}$O, pure CO$_{2}$, and the same ratio of gases as Earth's atmosphere. We also modeled atmospheres including 1 ppm, 100 ppm, or 1~\% of either H$_{2}$O or CO$_{2}$, with the rest of the atmosphere filled with O$_{2}$ as a stand-in for a background gas with no spectral features observable at MIRI/LRS wavelengths. Finally, we modeled the secondary eclipse depth for bare rock planets of the surface types described in \citet{Hu2012}.

\begin{figure*}
    \centering
    \includegraphics[width=\linewidth]{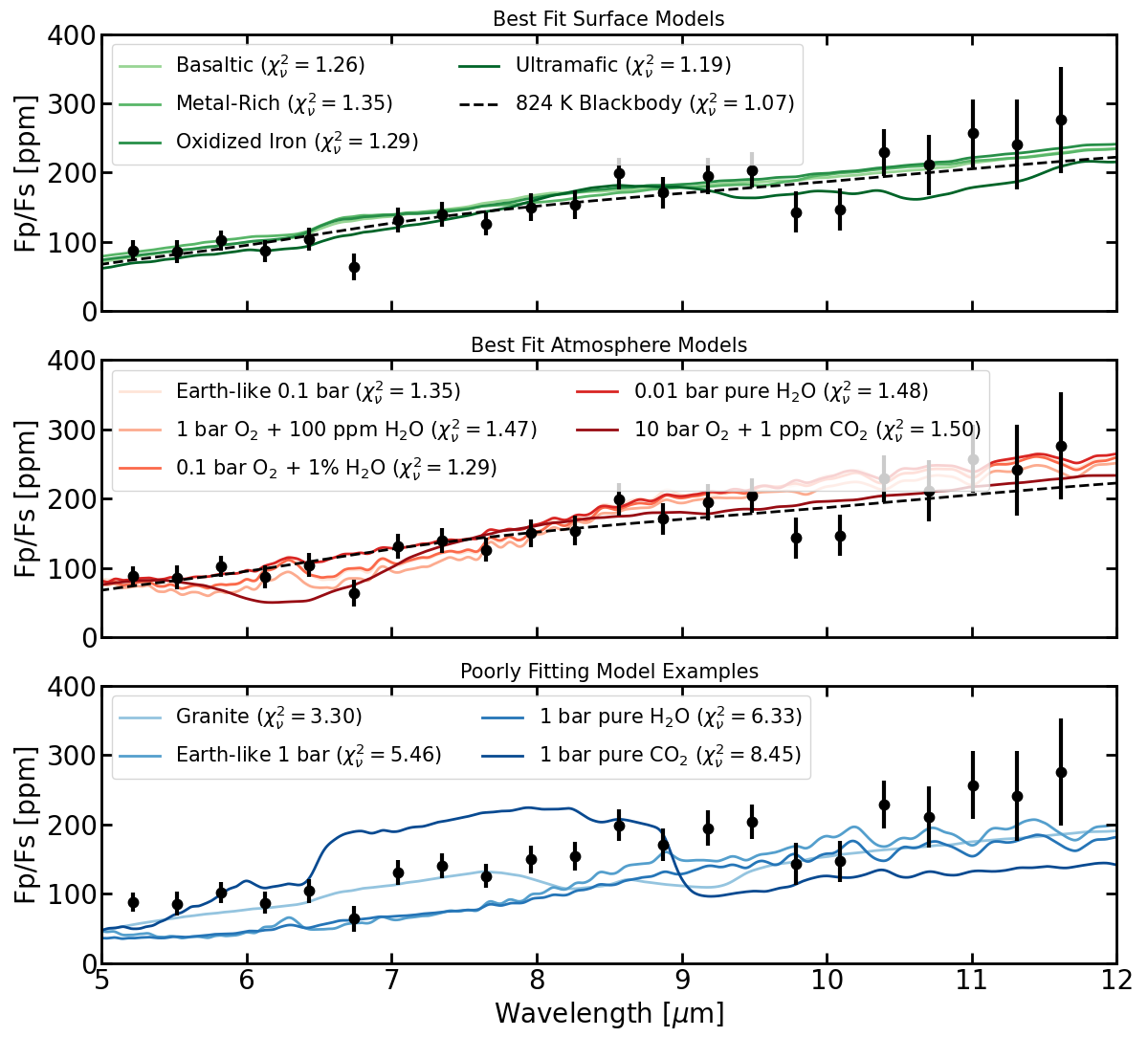}
    \caption{Comparison of the eclipse spectrum of Gl~486b (black points) with a variety of atmosphere and bare rock forward models. Top and middle panels show all bare rock and atmosphere models that have a reduced chi squared of $\leq 1.5$, respectively, assuming the number of degrees of freedom equals the number of data points. The bottom panel shows a selection of bare rock and atmosphere models which give worse fits to the data. The only atmospheres as consistent with our data as the bare rock models have relatively low surface pressures and/or small concentrations of the infrared absorbers H$_{2}$O and CO$_{2}$.}
    \label{fig:spec_models}
\end{figure*}

Figure~\ref{fig:spec_models} shows the best fitting forward models compared to the observed eclipse spectrum. Reduced chi squared values for the models quoted in this figure are calculated only from taking into account the error bars on the observed eclipse depths, not errors on the stellar and orbital parameters which could change the overall eclipse depths in the forward models themselves. Low-albedo bare rock surfaces (basalt, oxidized iron, ultramafic, or metal-rich rocks) are relatively consistent with the observed spectrum, with reduced chi squared values between $1.19 \leq \chi^{2}_{\nu} \leq 1.35$. All other forward models with $\chi^{2}_{\nu} \leq 1.5$ are shown in Figure~\ref{fig:spec_models}. The only atmosphere models whose fits are comparable in reduced chi squared with the bare rock models are those with small amounts of H$_{2}$O or CO$_{2}$ in relatively thin atmospheres. Figure~\ref{fig:spec_models} also shows the best-fit blackbody to the data, which had a dayside temperature of $824\pm12$~K and achieved a reduced chi-squared of 1.07.

The only O$_{2}$-rich atmospheres which were consistent with the data to within a reduced chi squared of 4 had surface pressures of $1-10$~bar and $<100$~ppm H$_{2}$O or CO$_{2}$ in them. Additionally, in order to have a reduced chi squared of less than 4, a pure H$_{2}$O, pure CO$_{2}$, or Earth-like atmosphere would have to have a surface pressure of $<0.1$~bar, $<10^{-4}$~bar, and $<0.01$~bar, respectively. The data strongly disfavor thicker atmospheres and/or those with larger concentrations of H$_{2}$O or CO$_{2}$.

\section{Discussion}
\label{sec:discuss}

As discussed in Section~\ref{sec:model}, our observed emission spectrum is most consistent with either airless bodies or atmospheres with modest surface pressures and small concentrations of infrared absorbers such as H$_{2}$O and CO$_{2}$. Our observations are highly inconsistent with atmospheres thicker than $\approx1$~bar with more than $\approx100$~ppm of H$_{2}$O or CO$_{2}$. \citet{Moran2023} observed Gl~486b in transit with \textit{JWST}/NIRSpec between $3-5$~$\mu$m. Their resulting transmission spectrum showed a slope toward the shorter wavelengths that could be described well by either a water-rich planetary atmosphere or an airless planet affected by unocculted starspots and the Transit Light Source Effect \citep{Rackham2018}. \citet{Moran2023} performed a retrieval for the water-rich case and found it preferred a composition with $>10$\% water. Our eclipse spectrum is inconsistent with a $>10$\% water atmospheric composition, regardless of the exact composition of the remaining 90\% of the atmosphere, because such an atmosphere would be expected to have much more prominent water absorption features in the MIRI/LRS bandpass.

\citet{DiamondLowe2024} observed the host star Gl~486b at X-ray and UV wavelengths and found multiple high-energy flares in a short time period. This result suggests that, although Gl~486 presents as relatively inactive at optical wavelengths, it may nonetheless be active enough in the UV to efficiently strip Gl~486b of any atmosphere. Given the inconsistency in water mass fraction for a potential atmosphere between our emission spectrum and the transit observations of \citet{Moran2023}, and the potential for Gl~486's activity to strip thin atmospheres over its $\geq6.6$-Gyr lifetime \citep{DiamondLowe2024}, we find that our observations are more consistent with an airless planet.

\begin{figure}
    \centering
    \includegraphics[width=\linewidth]{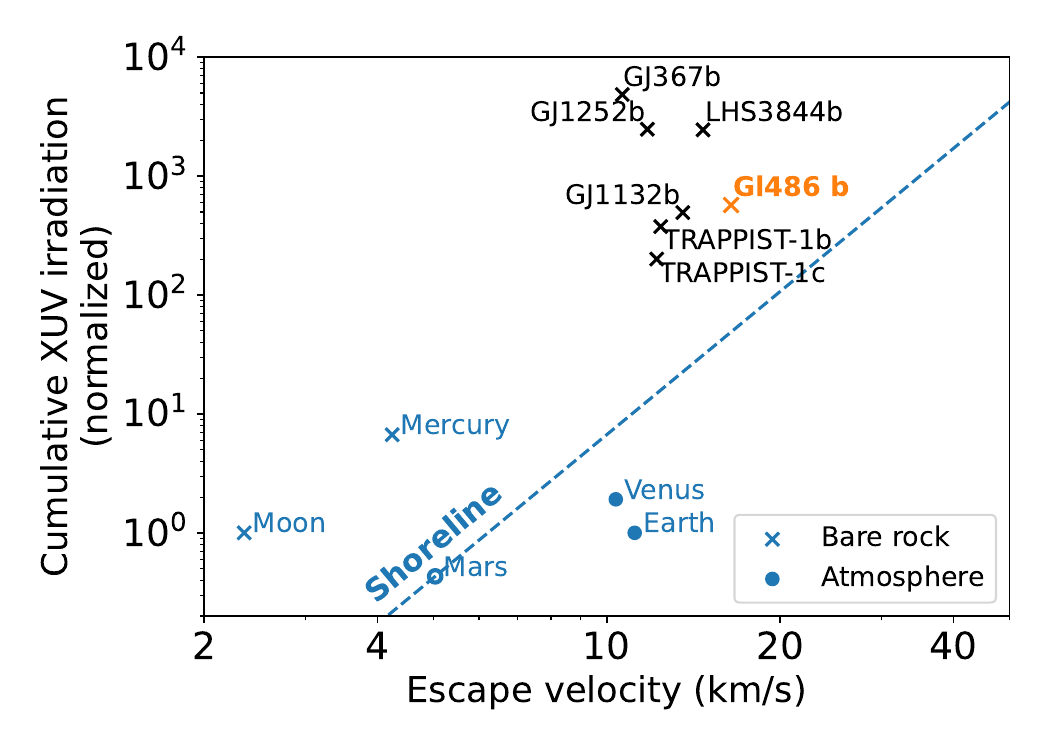}
    \caption{The Cosmic Shoreline \citep[dashed blue line;][]{Zahnle2017}, compared to select rocky bodies in the Solar System (blue) and in M Dwarf systems (black, orange). The plotted Cosmic Shoreline is a relationship between escape velocity and normalized cumulative XUV irradiation that is empirically observed in the Solar System. For exoplanets, crosses indicate planets whose observations rule out thick atmospheres and are consistent with no atmosphere. For Solar System bodies, crosses, open circles, and filled circles indicate bodies with no atmosphere, thin ($P<1$~bar) atmospheres, and thick ($P\geq 1$~bar) atmospheres, respectively. Gl 486b is 
    the second-closest exoplanet to the shoreline probed for an atmosphere so far, after TRAPPIST-1c.}
    \label{fig:shoreline}
\end{figure}

Our observation of Gl~486b's thermal emission resulted in the most precise constraint yet on the dayside temperature of a terrestrial planet (for example, \textit{JWST}/MIRI observations of TRAPPIST-1b and \textit{Spitzer} observations of LHS~3844b had $1\sigma$ dayside temperature errors of 26~K and 40~K, respectively). Therefore, although we did not detect strong evidence for an atmosphere, we can use this precise measurement to constrain the position of the Cosmic Shoreline for M-Earths. Figure~\ref{fig:shoreline} plots the escape velocity and cumulative XUV flux for the solar system terrestrial planets and the population of M-Earths observed with \textit{JWST} \citep{Greene2023,Zieba2023,zhang_2024,Xue2024} or \textit{Spitzer} \citep{Kreidberg2019,Crossfield2022}, assuming the scaling relationship for cumulative XUV flux from \citet{Zahnle2017}. The blue line shows the Cosmic Shoreline assuming that Mars, which retains a very thin atmosphere, resides exactly along the shoreline \citep{Zahnle2017}. Theory predicts that bodies to the upper left of this dividing line will lose their atmospheres to XUV-driven escape, while those to the lower right will retain their atmospheres. 

Our result suggests that future observations of M-Earths may need to probe closer to the Cosmic Shoreline to successfully detect thick atmospheres. However, the current sample of observed M-Earths is relatively small and heavily biased towards highly irradiated objects. Future surveys, such as the 500 hour DDT program to observe M-Earths with \textit{JWST} \citep{Redfield2024}, will increase the size of this sample and improve our understanding of the Cosmic Shoreline for M-Earths.


\section{Conclusions}
\label{sec:conclude}

We observed two secondary eclipses of the hot terrestrial exoplanet Gl~486b with \textit{JWST}/MIRI+LRS between $5-12$~$\mu$m. We jointly modeled these observations with previous \textit{JWST}/NIRSpec+G395H and \textit{TESS} transits to derive precise constraints on the stellar and planetary parameters. Based on these constraints, we derived a broadband dayside temperature of $T_{p}=865 \pm 14$~K and a temperature ratio of $\mathcal{R}=\frac{T_{p,dayside}}{T_{p,max}}=0.97 \pm 0.01$. While this value could be consistent with either an airless body with a small, non-zero albedo (i.e., the Moon or Mercury) or a thin atmosphere, it is inconsistent with Earth- or Venus-like atmospheres. In light of the previous transit observations of Gl~486b with \textit{JWST}/NIRSpec \citep{Moran2023} and the likelihood of a thin atmosphere escaping over the $\geq6.6$-Gyr lifetime of Gl~486 \citep{DiamondLowe2024}, we conclude that Gl~486b is most likely an airless rock. However, our measurement of the dayside temperature is the most precise of any \textit{JWST} thermal emission observations of terrestrial planets, which places a strong constraint on the position of the Cosmic Shoreline. This tight constraint can provide guidance for future searches for terrestrial exoplanet atmospheres with \textit{JWST}.


\begin{acknowledgments}
This work is based on observations with the NASA/ESA/CSA JWST. The data were obtained from the Mikulski Archive for Space Telescopes at the Space Telescope Science Institute (STScI), which is operated by the Association of Universities for Research in Astronomy, Incorporated, under NASA contract NAS5-03127. These observations are associated with program \#1743. Support for this program was provided through a grant from STScI. M.W.M. acknowledges support through the NASA Hubble Fellowship grant HST-HF2-51485.001-A awarded by STScI.  M.Z. acknowledges support through the 51 Pegasi b fellowship program awarded by the Heising-Simons Foundation. J. I. and E. M.-R. K. acknowledge support from the AEThER Matter-to-Life program, funded by the Alfred P. Sloan Foundation under grant G202114194. Work by A.S.M. and J.D.E. was funded by NASA ADAP 80NSSC19K1014. We thank the anonymous reviewer for their helpful comments.
\end{acknowledgments}

\vspace{5mm}
\facilities{JWST/MIRI}

\software{Eureka! \citep{Bell2022}, SPARTA \citep{Kempton2023}, batman \citep{batman}, astropy \citep{Astropy2013,Astropy2018,Astropy2022}, matplotlib \citep{Hunter2007}, numpy \citep{Harris2020}, pysynphot \citep{pysynphot}, scipy \citep{Virtanen2021}, ExoFASTv2 \citep{Eastman2019}, HELIOS \citep{Malik2017,Malik2019a,Malik2019b,Whittaker2022}}

\appendix 

\section{Stellar Parameters}
\label{append:star}

\startlongtable
\begin{deluxetable*}{lcccccccc}
\tablecaption{Median values and 68\% confidence interval for the \texttt{EXOFASTv2} fit to \textit{JWST} and \textit{TESS} observations of Gl~486b.}
\tablehead{\colhead{~~~Parameter} & \colhead{Description} & \multicolumn{7}{c}{Values}}
\startdata
\smallskip\\\multicolumn{3}{l}{Stellar Parameters:}\smallskip\\
~~~~$M_*$\dotfill &Mass (\msun)\dotfill &$0.3120^{+0.0070}_{-0.0069}$\\
~~~~$R_*$\dotfill &Radius (\rsun)\dotfill &$0.3243^{+0.0044}_{-0.0034}$\\
~~~~$R_{*,SED}$\dotfill &Radius$^{1}$ (\rsun)\dotfill &$0.3573^{+0.0054}_{-0.0061}$\\
~~~~$L_*$\dotfill &Luminosity (\lsun)\dotfill &$0.01151^{+0.00047}_{-0.00046}$\\
~~~~$F_{Bol}$\dotfill &Bolometric Flux (cgs)\dotfill &$5.64\pm0.23 \times 10^{-9}$\\
~~~~$\rho_*$\dotfill &Density (cgs)\dotfill &$12.95^{+0.25}_{-0.50}$\\
~~~~$\log{g}$\dotfill &Surface gravity (cgs)\dotfill &$4.9111^{+0.0068}_{-0.011}$\\
~~~~$T_{\rm eff}$\dotfill &Effective temperature (K)\dotfill &$3317^{+36}_{-37}$\\
~~~~$T_{\rm eff,SED}$\dotfill &Effective temperature$^{1}$ (K)\dotfill &$3197^{+23}_{-21}$\\
~~~~$[{\rm Fe/H}]$\dotfill &Metallicity (dex)\dotfill &$-0.15^{+0.13}_{-0.12}$\\
~~~~$K_S$\dotfill &Absolute Ks-band mag (mag)\dotfill &$6.823\pm0.020$\\
~~~~$k_S$\dotfill &Apparent Ks-band mag (mag)\dotfill &$6.359\pm0.020$\\
~~~~$A_V$\dotfill &V-band extinction (mag)\dotfill &$0.034^{+0.029}_{-0.024}$\\
~~~~$\sigma_{SED}$\dotfill &SED photometry error scaling \dotfill &$1.56^{+0.63}_{-0.37}$\\
~~~~$\varpi$\dotfill &Parallax (mas)\dotfill &$123.830\pm0.034$\\
~~~~$d$\dotfill &Distance (pc)\dotfill &$8.0756\pm0.0022$\\
\smallskip\\\multicolumn{2}{l}{Planetary Parameters:}&b\smallskip\\
~~~~$P$\dotfill &Period (days)\dotfill &$1.46712127^{+0.00000031}_{-0.00000035}$\\
~~~~$R_P$\dotfill &Radius ($R_{\oplus}$)\dotfill &$1.289^{+0.019}_{-0.014}$\\
~~~~$M_P$\dotfill &Mass ($M_{\oplus}$)\dotfill &$2.770^{+0.076}_{-0.073}$\\
~~~~$T_C$\dotfill &Observed Time of conjunction$^{2}$ (\bjdtdb)\dotfill &$2459691.128107^{+0.000060}_{-0.000054}$\\
~~~~$T_C$\dotfill &Model Time of conjunction$^{2,3}$ (\tjdtdb)\dotfill &$2459691.128008^{+0.000060}_{-0.000054}$\\
~~~~$T_T$\dotfill &Model time of min proj sep$^{3,4,5}$ (\tjdtdb)\dotfill &$2459939.071503^{+0.000010}_{-0.000011}$\\
~~~~$T_0$\dotfill &Obs time of min proj sep$^{4,6,7}$ (\bjdtdb)\dotfill &$2459939.071602\pm0.000010$\\
~~~~$a$\dotfill &Semi-major axis (AU)\dotfill &$0.01714\pm0.00013$\\
~~~~$i$\dotfill &Inclination (Degrees)\dotfill &$89.39^{+0.41}_{-0.42}$\\
~~~~$e$\dotfill &Eccentricity \dotfill &$0.00086^{+0.0016}_{-0.00043}$\\
~~~~$\omega_*$\dotfill &Arg of periastron (Degrees)\dotfill &$14^{+62}_{-85}$\\
~~~~$\dot{\omega}_{\rm GR}$\dotfill &Computed GR precession ($^\circ$/century)\dotfill &$4.832\pm0.072$\\
~~~~$T_{\rm eq}$\dotfill &Equilibrium temp$^{8}$ (K)\dotfill &$696.3^{+7.2}_{-7.3}$\\
~~~~$\tau_{\rm circ}$\dotfill &Tidal circ timescale (Gyr)\dotfill &$2.87^{+0.13}_{-0.19}$\\
~~~~$K$\dotfill &RV semi-amplitude (m/s)\dotfill &$3.389^{+0.076}_{-0.075}$\\
~~~~$R_P/R_*$\dotfill &Radius of planet in stellar radii \dotfill &$0.037244^{+0.000059}_{-0.000056}$\\
~~~~$a/R_*$\dotfill &Semi-major axis in stellar radii \dotfill &$11.380^{+0.074}_{-0.15}$\\
~~~~$\delta$\dotfill &$\left(R_P/R_*\right)^2$ \dotfill &$0.0013871^{+0.0000044}_{-0.0000042}$\\
~~~~$\delta_{\rm MIRILRS}$\dotfill &Transit depth in MIRI/LRS (frac)\dotfill &$0.00194^{+0.0015}_{-0.00043}$\\
~~~~$\delta_{\rm NRS1}$\dotfill &Transit depth in NIRSpec/G395H/NRS1 (frac)\dotfill &$0.001443^{+0.000026}_{-0.000025}$\\
~~~~$\delta_{\rm NRS2}$\dotfill &Transit depth in NIRSpec/G395H/NRS2 (frac)\dotfill &$0.001518^{+0.000027}_{-0.000030}$\\
~~~~$\delta_{\rm TESS}$\dotfill &Transit depth in TESS (frac)\dotfill &$0.00155^{+0.00020}_{-0.00011}$\\
~~~~$\tau$\dotfill &In/egress transit duration (days)\dotfill &$0.001545^{+0.000043}_{-0.000020}$\\
~~~~$T_{14}$\dotfill &Total transit duration (days)\dotfill &$0.042327^{+0.000041}_{-0.000035}$\\
~~~~$T_{FWHM}$\dotfill &FWHM transit duration (days)\dotfill &$0.040775^{+0.000028}_{-0.000027}$\\
~~~~$b$\dotfill &Transit impact parameter \dotfill &$0.120^{+0.080}_{-0.081}$\\
~~~~$b_S$\dotfill &Eclipse impact parameter \dotfill &$0.120^{+0.080}_{-0.081}$\\
~~~~$\tau_S$\dotfill &In/egress eclipse duration (days)\dotfill &$0.001546^{+0.000043}_{-0.000021}$\\
~~~~$T_{S,14}$\dotfill &Total eclipse duration (days)\dotfill &$0.04234^{+0.00014}_{-0.00011}$\\
~~~~$T_{S,FWHM}$\dotfill &FWHM eclipse duration (days)\dotfill &$0.040785^{+0.00013}_{-0.000097}$\\
~~~~$\delta_{S,2.5\mu m}$\dotfill &BB eclipse depth at 2.5$\mu$m (ppm)\dotfill &$1.67\pm0.11$\\
~~~~$\delta_{S,5.0\mu m}$\dotfill &BB eclipse depth at 5.0$\mu$m (ppm)\dotfill &$31.21^{+0.97}_{-0.94}$\\
~~~~$\delta_{S,7.5\mu m}$\dotfill &BB eclipse depth at 7.5$\mu$m (ppm)\dotfill &$73.7^{+1.5}_{-1.4}$\\
~~~~$\rho_P$\dotfill &Density (cgs)\dotfill &$6.66^{+0.23}_{-0.29}$\\
~~~~$logg_P$\dotfill &Surface gravity (cgs)\dotfill &$3.194^{+0.012}_{-0.014}$\\
~~~~$\Theta$\dotfill &Safronov Number \dotfill &$0.00813\pm0.00021$\\
~~~~$\fave$\dotfill &Incident Flux (\fluxcgs)\dotfill &$0.0533^{+0.0023}_{-0.0022}$\\
~~~~$T_S$\dotfill &Observed Time of eclipse$^{2}$ (\bjdtdb)\dotfill &$2459690.39472\pm0.00015$\\
~~~~$T_S$\dotfill &Model Time of eclipse$^{2,3}$ (\tjdtdb)\dotfill &$2459690.39482\pm0.00015$\\
~~~~$T_E$\dotfill &Model time of sec min proj sep$^{3,4,5}$ (\tjdtdb)\dotfill &$2460099.72165\pm0.00012$\\
~~~~$T_{E,0}$\dotfill &Obs time of sec min proj sep$^{4,6,7}$ (\bjdtdb)\dotfill &$2460099.72155\pm0.00012$\\
~~~~$T_P$\dotfill &Time of Periastron (\tjdtdb)\dotfill &$2459690.82^{+0.25}_{-0.35}$\\
~~~~$T_A$\dotfill &Time of asc node (\tjdtdb)\dotfill &$2459692.22857^{+0.00078}_{-0.00054}$\\
~~~~$T_D$\dotfill &Time of desc node (\tjdtdb)\dotfill &$2459691.49491^{+0.00057}_{-0.00072}$\\
~~~~$V_c/V_e$\dotfill &Scaled velocity \dotfill &$0.9999^{+0.0011}_{-0.0016}$\\
~~~~$e\cos{\omega_*}$\dotfill & \dotfill &$0.00040\pm0.00013$\\
~~~~$e\sin{\omega_*}$\dotfill & \dotfill &$0.0001^{+0.0016}_{-0.0011}$\\
~~~~$M_P\sin i$\dotfill &Minimum mass (\mj)\dotfill &$0.00871^{+0.00024}_{-0.00023}$\\
~~~~$M_P/M_*$\dotfill &Mass ratio \dotfill &$2.667^{+0.063}_{-0.062} \times 10^{-5}$\\
~~~~$d/R_*$\dotfill &Separation at mid transit \dotfill &$11.374^{+0.082}_{-0.15}$\\
~~~~$P_T$\dotfill &A priori non-grazing transit prob \dotfill &$0.08465^{+0.0011}_{-0.00060}$\\
~~~~$P_{T,G}$\dotfill &A priori transit prob \dotfill &$0.09119^{+0.0012}_{-0.00065}$\\
~~~~$P_S$\dotfill &A priori non-grazing eclipse prob \dotfill &$0.08457^{+0.0011}_{-0.00053}$\\
~~~~$P_{S,G}$\dotfill &A priori eclipse prob \dotfill &$0.09111^{+0.0012}_{-0.00057}$\\
\smallskip\\\multicolumn{3}{l}{Wavelength Parameters:}\smallskip\\
\multicolumn{3}{l}{MIRI/LRS}\\
 ~~~~$u_{1}$\dotfill &Linear limb-darkening coeff \dotfill &$0.58^{+0.61}_{-0.42}$\\
 ~~~~$u_{2}$\dotfill &Quadratic limb-darkening coeff \dotfill & $0.00^{+0.44}_{-0.43}$ \\
 ~~~~$\delta_{S}$\dotfill &Measured eclipse depth (ppm)\dotfill &$133.3^{+5.0}_{-4.9}$\\
 \multicolumn{3}{l}{NIRSpec/G395H/NRS1}\\
  ~~~~$u_{1}$\dotfill &Linear limb-darkening coeff \dotfill &$0.079^{+0.035}_{-0.034}$\\
  ~~~~$u_{2}$\dotfill &Quadratic limb-darkening coeff \dotfill & $0.149\pm0.053$ \\
 \multicolumn{3}{l}{NIRSpec/G395H/NRS2}\\
  ~~~~$u_{1}$\dotfill &Linear limb-darkening coeff \dotfill &$0.175^{+0.033}_{-0.037}$\\
  ~~~~$u_{2}$\dotfill &Quadratic limb-darkening coeff \dotfill & $-0.033^{+0.055}_{-0.048}$ \\
 \multicolumn{3}{l}{TESS}\\
  ~~~~$u_{1}$\dotfill &Linear limb-darkening coeff \dotfill &$0.21^{+0.21}_{-0.15}$\\~~~~$u_{2}$\dotfill &Quadratic limb-darkening coeff \dotfill & $0.34^{+0.25}_{-0.31}$ \\
\smallskip\\\multicolumn{3}{l}{Telescope Parameters:}\smallskip\\
\multicolumn{3}{l}{CARMENES}\\
~~~~$\gamma_{\rm rel}$\dotfill &Relative RV Offset (m/s)\dotfill &$-0.07\pm0.24$\\
~~~~$\sigma_J$\dotfill &RV Jitter (m/s)\dotfill &$1.72^{+0.22}_{-0.20}$\\
~~~~$\sigma_J^2$\dotfill &RV Jitter Variance \dotfill &$2.95^{+0.82}_{-0.66}$\\
\multicolumn{3}{l}{MAROON-X-Blue}\\
~~~~$\gamma_{\rm rel}$\dotfill &Relative RV Offset (m/s)\dotfill &$0.01\pm0.11$\\
~~~~$\sigma_J$\dotfill &RV Jitter (m/s)\dotfill &$0.61\pm0.13$\\
~~~~$\sigma_J^2$\dotfill &RV Jitter Variance \dotfill &$0.37^{+0.18}_{-0.14}$\\
\multicolumn{3}{l}{MAROON-X-Red}\\
 ~~~~$\gamma_{\rm rel}$\dotfill &Relative RV Offset (m/s)\dotfill &$0.039\pm0.059$\\
 ~~~~$\sigma_J$\dotfill &RV Jitter (m/s)\dotfill &$0.392^{+0.056}_{-0.052}$\\
 ~~~~$\sigma_J^2$\dotfill &RV Jitter Variance \dotfill &$0.153^{+0.047}_{-0.038}$\\
 \smallskip\\\multicolumn{3}{l}{Transit Parameters:}\smallskip\\
 \multicolumn{3}{l}{TESS UT 2020-03-21 (TESS)} \\
  ~~~~$\sigma^{2}$\dotfill &Added Variance \dotfill &$-2.609^{+0.058}_{-0.057} \times 10^{-7}$\\
  ~~~~$F_0$\dotfill &Baseline flux \dotfill &$1.0000385\pm0.0000060$\\
 \multicolumn{3}{l}{JWST UT 2022-12-25 (NIRSpec/G395H/NRS1)} \\
  ~~~~$\sigma^{2}$\dotfill &Added Variance \dotfill &$-2.4^{+5.1}_{-4.9} \times 10^{-10}$\\
  ~~~~$\tau_{Ramp}$\dotfill &Exponential Ramp \dotfill &$0.1321^{+0.0068}_{-0.0062}$\\
  ~~~~$A_{Ramp}$\dotfill &Amp of exp ramp \dotfill &$0.002000^{+0.000055}_{-0.000050}$\\
  ~~~~$F_0$\dotfill &Baseline flux \dotfill &$1.0011938^{+0.0000088}_{-0.0000087}$\\
 \multicolumn{3}{l}{JWST UT 2022-12-25 (NIRSPec/G395H/NRS2)} \\
  ~~~~$\sigma^{2}$\dotfill &Added Variance \dotfill &$3.6^{+7.4}_{-7.1} \times 10^{-10}$ \\
  ~~~~$\tau_{Ramp}$\dotfill &Exponential Ramp \dotfill &$0.0374^{+0.012}_{-0.0077}$\\
  ~~~~$A_{Ramp}$\dotfill &Amp of exp ramp \dotfill &$9.5\pm1.2 \times 10^{-5}$\\
  ~~~~$F_0$\dotfill &Baseline flux \dotfill &$1.000460\pm0.000012$\\
 \multicolumn{3}{l}{JWST UT 2022-12-29 (NIRSpec/G395H/NRS1)} \\
  ~~~~$\sigma^{2}$\dotfill &Added Variance \dotfill &$-0.4^{+4.7}_{-4.5} \times 10^{-10}$\\
  ~~~~$\tau_{Ramp}$\dotfill &Exponential Ramp \dotfill &$0.194^{+0.014}_{-0.013}$\\
  ~~~~$A_{Ramp}$\dotfill &Amp of exp ramp \dotfill &$0.00253^{+0.00012}_{-0.00011}$\\
  ~~~~$F_0$\dotfill &Baseline flux \dotfill &$1.0011544^{+0.0000082}_{-0.0000081}$\\
 \multicolumn{3}{l}{JWST UT 2022-12-29 (NIRSpec/G395H/NRS2)} \\
  ~~~~$\sigma^{2}$\dotfill &Added Variance \dotfill &$0.1^{+6.1}_{-5.9} \times 10^{-10}$\\
  ~~~~$\tau_{Ramp}$\dotfill &Exponential Ramp \dotfill &$0.044^{+0.63}_{-0.021}$\\
  ~~~~$A_{Ramp}$\dotfill &Amp of exp ramp \dotfill &$-5.5^{+1.6}_{-7.3} \times 10^{-5}$\\
  ~~~~$F_0$\dotfill &Baseline flux \dotfill &$1.000361^{+0.000014}_{-0.000017}$\\
 \multicolumn{3}{l}{JWST UT 2023-05-29 (MIRI/LRS)} \\
  ~~~~$\sigma^{2}$\dotfill &Added Variance \dotfill &$-1.2^{+9.0}_{-8.8} \times 10^{-10}$\\
  ~~~~$\tau_{Ramp}$\dotfill &Exponential Ramp \dotfill &$0.0151^{+0.011}_{-0.0044}$\\
  ~~~~$A_{Ramp}$\dotfill &Amp of exp ramp \dotfill &$9.7^{+2.5}_{-1.8} \times 10^{-5}$\\
  ~~~~$F_0$\dotfill &Baseline flux \dotfill &$1.000094^{+0.000020}_{-0.000026}$\\
  ~~~~$C_{0}$\dotfill &Additive detrending coeff \dotfill &$-0.000555\pm0.000011$\\
   ~~~~$C_{1}$\dotfill &Additive detrending coeff \dotfill &$-5.8\pm1.2 \times 10^{-5}$\\
 \multicolumn{3}{l}{JWST UT 2023-06-01 (MIRI/LRS)}\\
  ~~~~$\sigma^{2}$\dotfill &Added Variance \dotfill &$-2.1^{+9.1}_{-8.9} \times 10^{-10}$\\
 ~~~~$\tau_{Ramp}$\dotfill &Exponential Ramp \dotfill &$0.35^{+0.45}_{-0.24}$\\
 ~~~~$A_{Ramp}$\dotfill &Amp of exp ramp \dotfill &$0.00039^{+0.00039}_{-0.00020}$\\
 ~~~~$F_0$\dotfill &Baseline flux \dotfill &$1.0000686^{+0.000012}_{-0.0000080}$\\
 ~~~~$C_{0}$\dotfill &Additive detrending coeff \dotfill &$-0.0004216^{+0.000010}_{-0.0000100}$\\
 ~~~~$C_{1}$\dotfill &Additive detrending coeff \dotfill &$-3.7\pm1.1 \times 10^{-5}$\\
\enddata
\label{tab:GJ486_stellar}
\tablenotetext{}{See Table 3 in \citet{Eastman2019} for a detailed description of all parameters}
\tablenotetext{1}{This value ignores the systematic error and is for reference only}
\tablenotetext{2}{Time of conjunction is commonly reported as the ``transit time''}
\tablenotetext{3}{\tjdtdb is the target's barycentric frame and corrects for light travel time}
\tablenotetext{4}{Time of minimum projected separation is a more correct ``transit time''}
\tablenotetext{5}{Use this to model TTVs, e}
\tablenotetext{6}{At the epoch that minimizes the covariance between $T_C$ and Period}
\tablenotetext{7}{Use this to predict future transit times}
\tablenotetext{8}{Assumes no albedo and perfect redistribution}
\end{deluxetable*}

\bibliography{main}{}
\bibliographystyle{aasjournal}

\end{document}